\documentclass[aip,graphicx,cha,floatfix]{revtex4-1}

\usepackage{graphicx}
\usepackage{amsmath}
\usepackage{amssymb}
\usepackage{natbib}
\usepackage{color}

\begin{document}


\title{Tunable power law in the desynchronization events of coupled chaotic electronic circuits} 




\author{Gilson F. \surname{de Oliveira Jr.}}
\email[corresponding author: ]{gilson@otica.ufpb.br}
\affiliation{Departamento de F\'{\i}sica, Universidade Federal da Para\'{\i}ba, 58051-900, Jo\~{a}o Pessoa, PB, Brazil.}

\author{Hugo L. D. de Souza \surname{Cavalcante}}
\affiliation{Departamento de Inform\'{a}tica, Universidade Federal da Para\'{i}ba, 58051-900, Jo\~{a}o Pessoa, PB, Brazil.}


\author{Orlando di Lorenzo}
\affiliation{Departamento de F\'{\i}sica, Universidade Federal da Para\'{\i}ba, 58051-900, Jo\~{a}o Pessoa, PB, Brazil.}

\author{Martine Chevrollier}
\affiliation{Departamento de F\'{\i}sica, Universidade Federal da Para\'{\i}ba, 58051-900, Jo\~{a}o Pessoa, PB, Brazil.}

\author{Thierry Passerat de Silans}
\affiliation{Departamento de F\'{\i}sica, Universidade Federal da Para\'{\i}ba, 58051-900, Jo\~{a}o Pessoa, PB, Brazil.}

\author{Marcos Ori\'{a}}
\affiliation{Departamento de F\'{\i}sica, Universidade Federal da Para\'{\i}ba, 58051-900, Jo\~{a}o Pessoa, PB, Brazil.}


\date{\today}

\begin{abstract}
We study the statistics of the amplitude of the synchronization error in chaotic electronic circuits coupled through linear feedback. 
Depending on the coupling strength, our system exhibits three qualitatively different regimes of synchronization: weak coupling yields independent oscillations; moderate to strong coupling produces a regime of intermittent synchronization known as attractor bubbling; and stronger coupling produces complete synchronization. 
In the regime of moderate coupling, the probability distribution for the sizes of desynchronization events follows a power law, with an exponent that can be adjusted by changing the coupling strength. Such power-law distributions are interesting, as they appear in many complex systems. However, most of the systems with such a behavior have a fixed value for the exponent of the power law, while here we present an example of a system where the exponent of the power law is easily tuned in real time. 
\end{abstract}


\maketitle 


\begin{quotation}
Since the discovery that chaotic systems may synchronize their trajectories in spite of the sensitive dependence to perturbations in each system\cite{Pecora1990}, the subject of chaos synchronization has attracted the attention of many researchers, both from the fundamental and applied point of view, and many different forms of synchronization have been reported under diverse conditions. 
Likewise, the statistics of large events whose size-distribution follows power laws  or other heavy-tailed models in complex systems has attracted the attention of multidisciplinary research. One of the difficulties arising in the study of many complex systems is the lack of reproducibility of the experiments under controlled conditions, issuing from the very complex nature of these usually large systems.  
Here we use imperfect chaos synchronization to generate rare events and further develop an analogy between heavy-tailed distributions appearing in complex systems and the statistics of desynchronization events in simple coupled chaotic systems\cite{Cavalcante2013}. 
\end{quotation}

\section{Introduction}
Synchronization phenomena are widely studied in many real and idealized systems, such as electronic circuits \cite{cuomo1993circuit, kim2006synchronization, senthilkumar2010experimental}, lasers \cite{zamora2010crowd, nixon2011synchronized, deshazer2001detecting}, and maps \cite{Maritan1994,Pyragas1996,Herzel1995,masoller2001delayed}. While one often studies the synchronization  of nonlinear dynamical systems in periodic states\cite{Acebron2005}, the synchronization of chaotic systems  is even more interesting, due to the counterintuitive effect of synchronization between two systems whose trajectories are exponentially sensitive to perturbations in phase-space\cite{boccaletti2002synchronization}, and also due to potential applications, such as masked communications \cite{Pecora1990,cuomo1993circuit}. Besides complete synchronization, one can find other generalizations to the concept, such as phase-synchronization\cite{Rosenblum1996} and generalized synchronization\cite{Rulkov1995,Kocarev1996}. When subject to a generic coupling, most systems will wander off the synchronized state because of noise or mismatched parameters. This is often the case of generic feedback-coupled nonlinear oscillators in which there are riddled basins of attraction\cite{Alexander1992,ashwin1994bubbling,Maistrenko1998}
that give rise to a phenomenon called attractor bubbling \cite{ashwin1994bubbling, heagy1995desynchronization, ashwin1996attractor, gauthier1996intermittent, venkataramani1996transitions, krawiecki2002volatility, flunkert2009bubbling,  krawiecki2009microscopic}.  As a consequence of attractor bubbling, the difference between the state variables, observed in the coupled elements (which we refer to as an error signal) shows long intervals of low values interspersed with sudden and brief departures to large values, which we call bubbles, bursts, or desynchronization events. 

The distribution of sizes of these bursts have characteristic parameters similar to the ones observed in extreme events that occur in complex systems \cite{Bak1999, Cavalcante2013}. This similarity suggests that we can use coupled chaotic oscillators as a proxy for the study of extreme events in complex systems, motivated by the problem of catastrophic behavior of many natural and artificial systems\cite{Bak1987, Bak1999, White1998, Solli2007, Bonatto2011}. Such complex system variables usually present non-normal statistical distributions with large values of event sizes associated to the asymptotic behavior of the distribution, which can be a power law or other heavy-tailed distribution\cite{estoup1916gammes, willis1922some, mercadier2009levy, newman2005power, sornette1998multiplicative}. The value of this slope can reveal information about the mechanism producing the bursts. Complex systems typically follow power laws $P(x) = C x^{-\alpha}$ with a characteristic exponent $\alpha$. For instance, $\alpha = 1.5$ in neurological systems \cite{Klaus2011}, or $\alpha = 2.0$, in coupled oscillators desynchronization \cite{Cavalcante2013}, or $\alpha = 1.0$ in earthquakes analysis \cite{Christensen2002}. Power-law distributions with exponents $\alpha > 3$ have finite first and second moments (average and variance, respectively), while for $3\geq \alpha > 1$ the second moment (variance) does not exist, and for $2\geq \alpha > 1$ even the first moment (average) is infinite. Here we show that, beyond the analysis of the probabilities of events of different sizes it is possible to tune the power law exponent itself, which has as a consequence the change of the value of the maximum event expected to occur during a given observation time. It is known that variables of many complex systems, including certain chaotic systems and systems on the edge of chaos\cite{Bak1999}, follow power law statistics. Understanding the physical mechanisms for the origin of a heavy-tailed distribution of event sizes  in a specific complex system may give important information about this system. For instance: power laws are scale-free, implying that large-, medium- and small-sized events all share a common mechanisms of formation. Such a lack of distinction  affects  the predictability of large events, as they are not discernible from the smaller ones in the earlier stages of formation. 

We analyze here a nonlinear dynamical system formed by a pair of coupled chaotic oscillators and we show that the statistical distribution of one of the variables describing the state of this system presents a heavy-tailed behavior, with a very interesting property: it can evolve through a large range of the power-law exponent, which can be tuned through a control parameter. Systems with a tunable-exponent power law do not occur very often in Nature, and finding one example of a simple and rather generic system with this property can give a hint on how the exponent could be controlled in other systems where it is important.  The system studied here is comprised of two electronic oscillators, each governed by a second-order differential equation driven by an external periodic force \cite{goncalves2011electrical}. We use the difference between the values of the same dynamical variable in the two oscillators to generate a unidirectional feedback, coupling the oscillators in a drive/response configuration. We call \emph{error signal} the difference between the variables of the oscillators, and we analyse its statistics for different values of the coupling strength. 
Our system exhibits three qualitatively different behaviors as we change the coupling strength: for strong coupling we observe perfect (or high-quality) synchronization; for weak or negligible coupling, the two subsystems are independent, while for intermediate ranges of coupling strength the distribution of burst sizes has a heavy tail, with large events becoming increasingly rare, as for example in power-law (Pareto) distributions found in L\'evy Flights \cite{Levy1937,Mandelbrot1982,Shlesinger1993}. 

\section{Circuit Description}
\begin{figure}[!htb]
\resizebox{8.5cm}{!}{\includegraphics{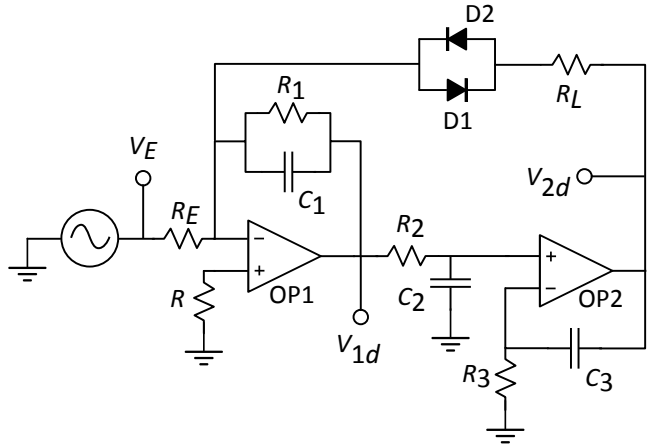}}
\caption{\label{fig:1circuit}Schematic diagram of the electronic circuit for the drive chaotic oscillator. This circuit implements a second-order differential equation for the voltage $V_{1d}$ driven by the external sinusoidal signal $V_E$. The small circles connected to wires indicate the points of measurement of $V_E$, $V_{1d}$ and its time integral $V_{2d}$.}
\end{figure}
The schematic diagram of the circuit that we used as one of the chaotic oscillators (the \emph{drive} system) is shown in Fig.\ \ref{fig:1circuit}. It is composed of commercial-grade resistors, capacitors, diodes and operational amplifiers. The values of the components in the drive circuit are: $R_{1}$ = 46.50 k${\Omega}$, $R = 14.86$ k${\Omega}$, $R_{2}$ = 14.85 k${\Omega}$, $R_{E}$ = 14.86 k${\Omega}$, $R_{L}$ = 512 ${\Omega}$, $R_{3}$ = 14.85 k${\Omega}$, $C_{1}$ = 14.73 nF, $C_{2}$ = 14.83 nF, $C_{3}$ = 14.83 nF. $D_{1}$ and $D_{2}$ are 2N4148 diodes and  the operational amplifiers OP1 and OP2 are LF411CN.

\begin{figure}[!htb]
\resizebox{8.0cm}{!}{\includegraphics{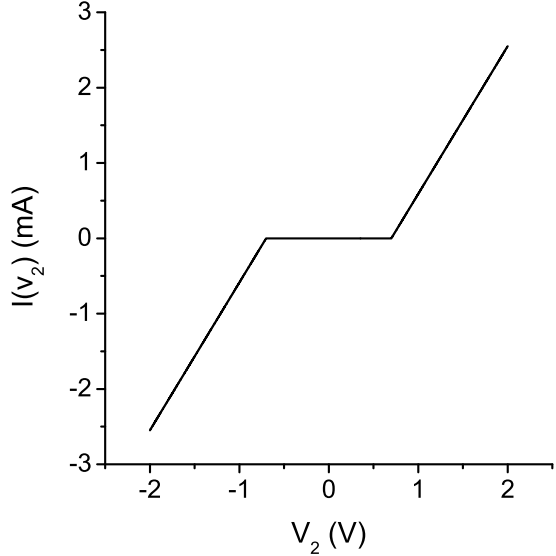}}
\caption{\label{fig:DiodeCurve}Nonlinear current through a pair of anti-parallel diodes, such as $D_1$ and $D_2$ in Fig.\ \ref{fig:1circuit}. This is the idealized piecewise-linear current given by Eq.\ (\ref{eq:current}). }
\end{figure}

Analysis of Kirchhoff's laws reveals that the dynamical state of the circuit can be expressed in terms of the voltages $V_{1d}$, $V_{2d}$ and $V_E(t)$, which obey a second-order differential equation with an external pumping when $R_2 = R_3$ and $C_2 = C_3$\cite{goncalves2011electrical}. This second-order differential equation can be written as two first-order equations:
\begin{eqnarray}
\dot{V}_{1d} & = & -\gamma V_{1d}
	-\alpha I(V_{2d})
	-\beta V_E, \label{eq:V1_dot} \\ 
\dot{V}_{2d} & = & \theta V_{1d} \label{eq:V2_dot},
\end{eqnarray}
where $\gamma = 1/(R_{1}C_{1})$, $\alpha = 1/C_{1}$, $\beta = 1/(R_{E}C_{1})$, $\theta = 1/(R_{2}C_{2})$, and the characteristic time scale of the evolution of our system is $1/\theta$, given by $R_2C_2 = 220.2\ \mu$s. $V_{1d}$ and $V_{2d}$ are  the voltages at the outputs of OP1 and OP2, respectively, $V_E = A \sin{2{\pi} f t}$ is the external pump voltage with amplitude $A$ and frequency $f$, and $I(V_{2d})$ is  the current passing through the diodes D1 and D2 and going to the inverting input of OP1, approximately given by 
\begin{equation}
I(V_{2d})  = \left\{\begin{array}{cr}
	(V_{2d}+0.7)/\text{R}_\text{L}, & \text{if}\ V_{2d} < -0.7,  \\
	0, & \text{if}\ -0.7 \leq V_{2d} \leq 0.7,  \\
	(V_{2d}-0.7)/\text{R}_\text{L}, & \text{if}\ V_{2d} > 0.7,  \\
	\end{array} \right. \label{eq:current}
\end{equation}
and plotted in Fig.\ \ref{fig:DiodeCurve}. 

Let us recall that, in order for a continuous, autonomous system to be chaotic its dynamics needs to i) be embedded in, at least,  a three-dimensional (3D) phase space \cite{Strogatz1994}, and ii) have a nonlinear term. Our system has a nonlinearity in the current of Eq.\ (\ref{eq:current}), and the phase of the external forcing can be recast as a dynamical variable that provides for the third dimension.

\begin{figure}[!htb]
\includegraphics[width=16cm,height=8cm]{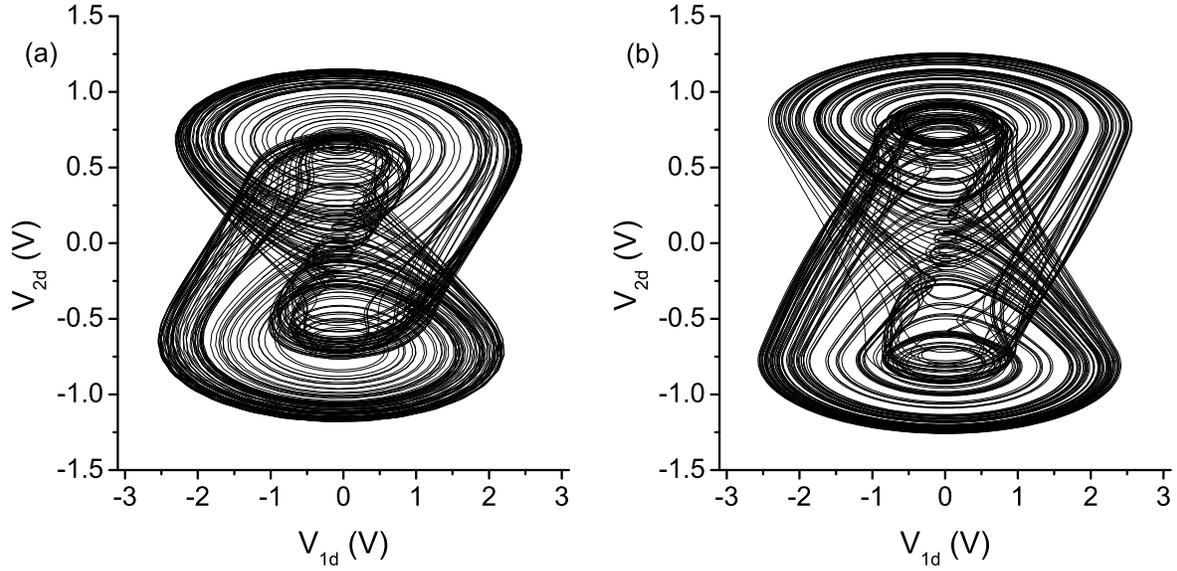}
\caption{\label{fig:ChaoticAttractor}A chaotic trajectory in the $V_{1d}$-$V_{2d}$ plane is plotted (a) from experimental data obtained with external pumping $A$ = 4.0 V and $f$ = 770 Hz and (b) from numerical data obtained through integration of Eqs. (\ref{eq:V1_dot}) and (\ref{eq:V2_dot}) using the second-order Runge-Kutta method, with external pumping $A$ = 3.0 V and $f$ = 720 Hz. The values of the other parameters are given in the text. }
\end{figure}

By changing the values of frequency and amplitude of the external drive one can tune the system to periodic or chaotic states, in both the experimental circuit and in the numerical model. 
In the experimental setup, the system exhibits windows of chaotic dynamical state, for example for frequencies between 760 Hz and 820 Hz and between 950 Hz and 1.0 kHz, both with amplitude 4.0 V.
Figure\ \ref{fig:ChaoticAttractor} shows a typical chaotic trajectory in phase space, acquired from the circuit (Fig.\ \ref{fig:ChaoticAttractor}(a)), and obtained by numerical integration of the differential equations (\ref{eq:V1_dot}) and (\ref{eq:V2_dot}) (Fig.\ \ref{fig:ChaoticAttractor}(b)).
As we use a simplified model to describe the circuit of Fig.\ \ref{fig:1circuit}, we need to do a few adjustments on the parameters amplitude and frequency of the external pump voltage $V_{E}$ in the simulation in order to obtain a dynamical behavior similar to the one observed in the experiment.
We are able to reproduce, in the model, all the observed oscillatory regimes, chaotic or periodic, with values of the parameters $f$ and $A$ slightly different from the ones used in the experiment. 
The values of resistors and capacitors are the same in the experiment and in the numerical integrations. 

\begin{figure} [!htb]
\resizebox{8.5cm}{!}{\includegraphics{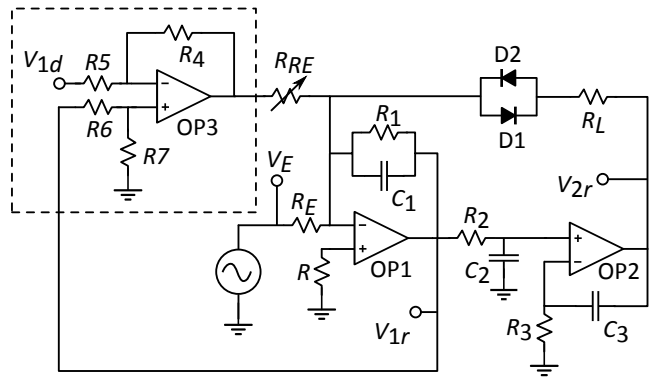}}
\caption{\label{fig:2circuits}Schematic diagram of the response system. The dashed rectangle highlights the feedback circuit that couples the response system to the drive system.}
\end{figure}

In order to couple two identical circuits, the voltage $V_1d$ of the drive circuit is injected, through a feedback circuit, into the second chaotic system. In Fig.\ \ref{fig:2circuits} we show the response circuit, highlighting the additional circuitry where the feedback signal is produced by a subtractor (operational-amplifier OP3) whose input voltages are $V_{1d}$ and $V_{1r}$ and the output voltage is $V_{1d}-V_{1r}$. This feedback  signal is added to the dynamics of the response subsystem, providing the coupling between the drive and the response. The parameter that measures the coupling level between the circuits is $\epsilon = R_{2}/R_{RE}$, where $R_{RE}$ is a variable resistor placed at the output of the subtractor.

The values of the components of the response circuit  are chosen to be as close as possible to their counterparts in the drive circuit, within a tolerance of 0.5\%. Therefore, the components $R_{1}$, $R$, $R_{2}$, $R_{E}$,  $R_{3}$, $C_{1}$, 
 have the same values in the drive and in the response circuits, while $C_{2}$ = 14.87 nF, $C_{3}$ = 14.87 nF, $R_{L}$ = 511 ${\Omega}$ have values measurably different in the response and drive circuits. The components of the subtractor circuit are: $R_{4}$ = 14.85 k${\Omega}$, $R_{5}$ = 14.85 k${\Omega}$, $R_{6}$ = 14.85 k${\Omega}$, $R_{7}$ = 14.85 k${\Omega}$. The response circuit counterparts of $D_{1}$, $D_{2}$, OP1 and OP2, are also the same models as in the drive circuit, and the operational amplifier OP3 is a LF411CN. A single signal generator provides the sinusoidal signal $V_E$ 
applied to both the drive and response systems.

The dynamic equations are the same for both circuits, except the coupling term, which is present only in the response circuit:
\begin{eqnarray}
\dot{V}_{1r} & = & -\gamma V_{1r}+
	\alpha I(V_{2r})
	-\beta V_{E} 
	+\frac{\epsilon}{R_{2}C_{1}} (V_{1d}-V_{1r}) \label{eq:V1_dot_response}, \\
\dot{V}_{2r} & = & \theta V_{1r} \label{eq:V2_dot_response}.
\end{eqnarray}

In the numerical integration of the coupled system we used amplitude $A =$ 3.0 V and frequency $f =$ 720 Hz in Eqs. (\ref{eq:V1_dot}), (\ref{eq:V2_dot}), (\ref{eq:V1_dot_response}) and (\ref{eq:V2_dot_response}), with a parameter mismatch $\sim$ 1\%, compound of the tolerances of the components. We set the parameters so that the experimental and numerical systems were both in the same chaotic state when uncoupled ($\epsilon = 0$).

\begin{figure}[!htb]
\resizebox{8.5cm}{!}{\includegraphics{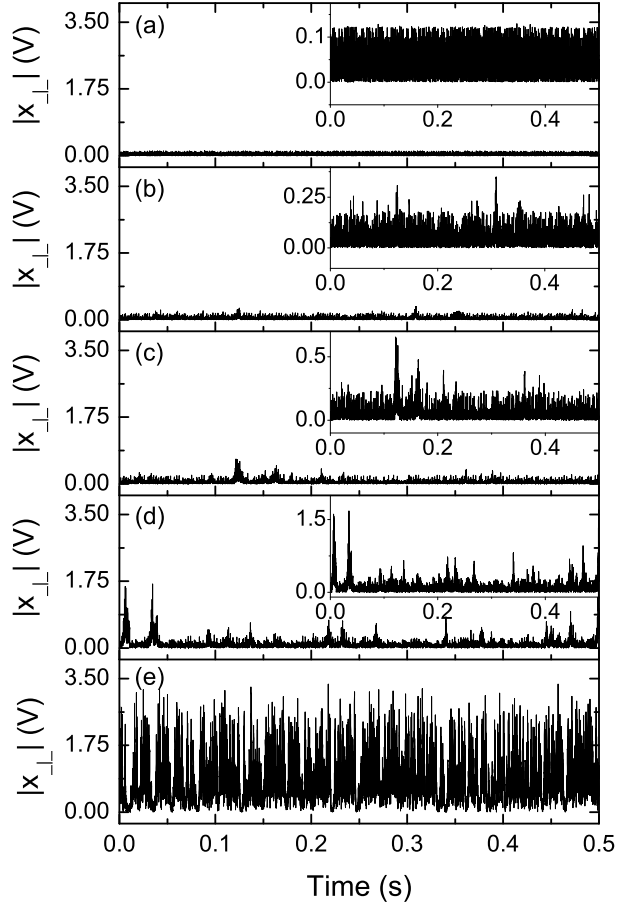}}
\caption{\label{fig:ErrorSignal}Short segments of the experimental time series of the error signal ($|x_{\bot}|=|V_{1d}-V_{1r}|+|V_{2d}-V_{2r}|$) for the coupled circuits with external pumping $A$ = 4.0 V and $f$ = 770 Hz, for different coupling levels: (a) $\epsilon$ = 1.0, (b) $\epsilon$ = 0.7, (c) $\epsilon$ = 0.6, (d) $\epsilon$ = 0.5, (e) $\epsilon$ = 0.0. The insets in (a), (b), (c) and (d) show the same time series in an amplified scale to exhibit details of the bursts in the error-signal time series. }
\end{figure}

In the experimental setup, the circuits were then coupled and we used a digital oscilloscope to acquire temporal series of the differences $V_{1d}-V_{1r}$  and $V_{2d}-V_{2r}$. From these signals we obtained the distance $|x_{\bot}|$ between the drive and the response systems in the 3D phase-space. In order to facilitate the analog calculation of the distance, we used the L-1 norm to define this distance as $|x_{\bot}|=|V_{1d}-V_{1r}|+|V_{2d}-V_{2r}|$ \footnote{In the L-1 norm the distance between two arbitrary points, $\vec{x}_a=(x_{a1},x_{a2},x_{a3})$ and $\vec{x}_b=(x_{b1},x_{b2},x_{b3})$, is given by $d = |x_{a1}-x_{b1}|+|x_{a2}-x_{b2}|+|x_{a3}-x_{b3}|$, instead of the square root of the sum of the squared differences, usual in the L-2 norm.}. Time series of $|x_{\bot}|$ are shown in Fig.\ \ref{fig:ErrorSignal} for different values of the coupling parameter $\epsilon$. Notice that the third dimension does not contribute to the distance, as both oscillators share the same value of $V_E$ (the full system is indeed 5D, instead of 6D). To compare the experimental and numerical results we used the same procedure in our integrations: we calculated the time series of $V_{1d}$, $V_{2d}$, $V_{1r}$ and $V_{2r}$ and then the distance (L-1 norm \cite{Note1}) between the drive and response systems $|x_{\bot}|$. The variable $|x_{\bot}|$ is then used to make the statistical analysis of the amplitudes of the dessynchronization bursts, discussed in the next section.

\section{Statistics of the error signal}
\label{sec:statistics}
\begin{figure}[htbp]
\resizebox{8.5cm}{!}{\includegraphics{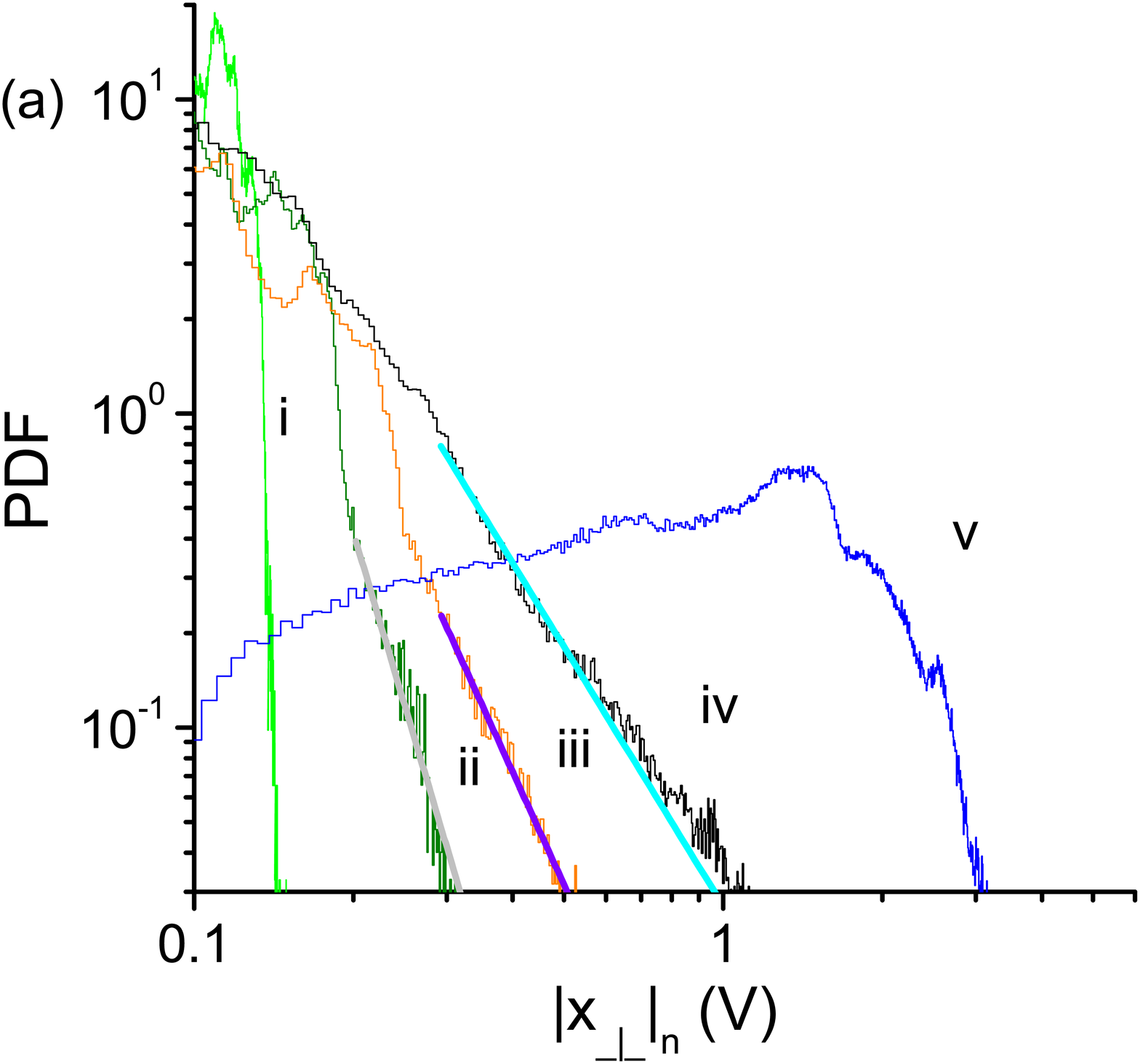}}
\resizebox{8.5cm}{!}{\includegraphics{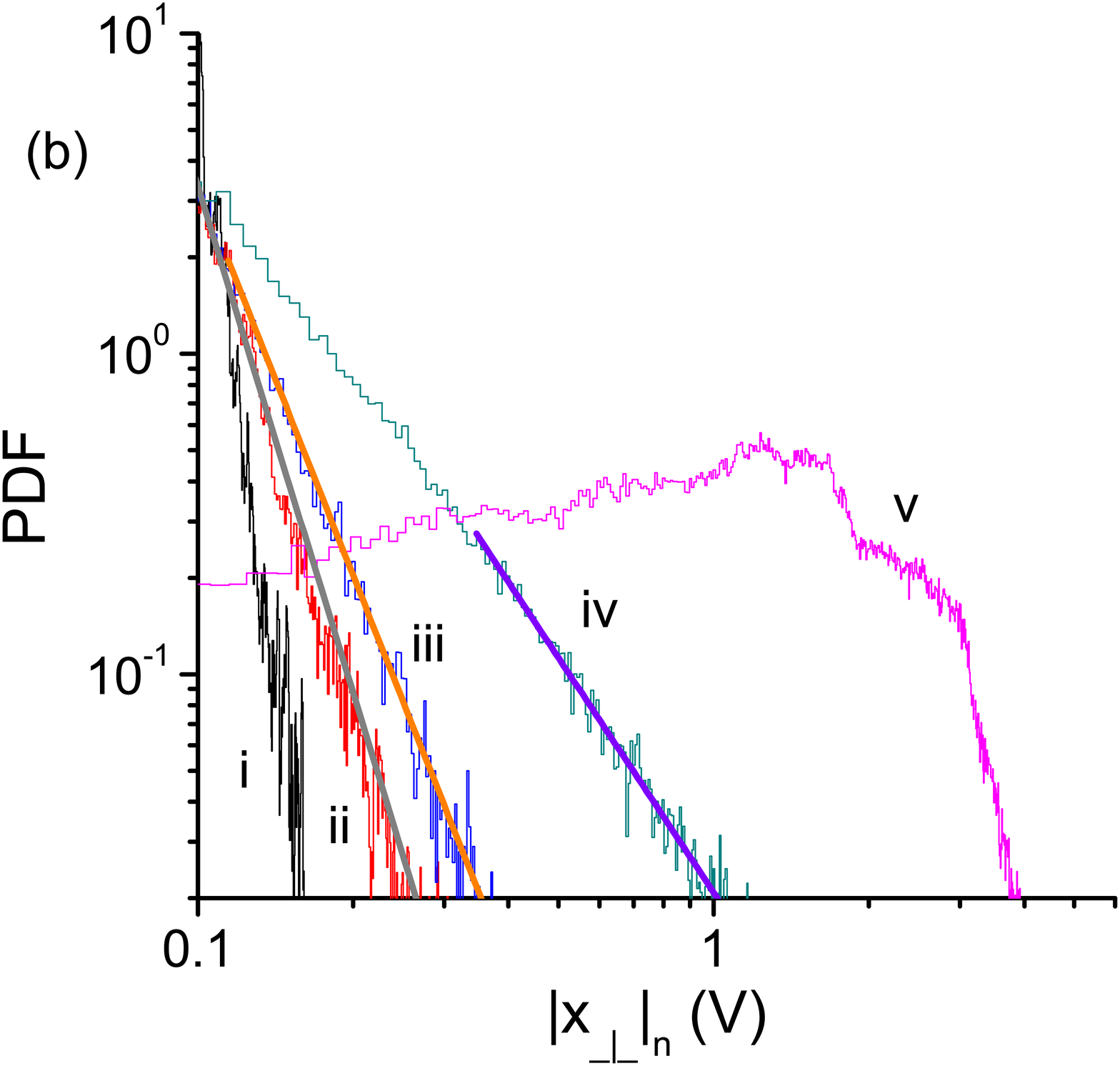}}
\caption{\label{fig:HistError}Log-log distribution of the voltages obtained from the maxima $|x_{\bot}|_n$ for different coupling levels $\epsilon$ with (i) $\epsilon$ = 1.0, (ii) $\epsilon$ = 0.7, (iii) $\epsilon$ = 0.6, (iv) $\epsilon$ = 0.5, (v) $\epsilon$ = 0.0: (a) experimental results  and (b) numerical results. The values of power-law exponents are shown in Table \ref{tab:exponents}.}
\end{figure}

 We acquired time series of  $|x_{\bot}|$ with 10$^7$ points at a sample rate of 100 kHz (sampling time 10 $\mu$s, series duration 100 s). To define a burst size, we first eliminate high-frequency fluctuations (noise) by applying a running average of 9 points to the time series of $|x_{\bot}|$ and build the temporal sequence of local maxima $|x_{\bot}|_n$ of the error signal. A maximum is detected by comparing a value of $|x_{\bot}|$, above a threshold of 0.05 V, with its 8 nearest neighbors on both sides. The running average prevents the detection of undesired high-frequency features in the error signal (false maxima caused by noise), but the running window is shorter than the characteristic time-scale of the dynamics, so that it does not average out pulses that originate from the dynamics. The threshold in the values of accepted maxima eliminates maxima that are too small and are below the lower cutoff of the  power law. The distributions of maxima are shown in Fig.\ \ref{fig:HistError} for different values of $\epsilon$. These empirical distributions are obtained as normalized histograms that have unit area in a linear scale. The experimental and numerical maxima distributions present the same general behavior, as we can see in Figs.\ \ref{fig:HistError}(a) and \ref{fig:HistError}(b). The histograms show a  qualitative change between three regimes when the coupling level is varied from 0.0 to 1.0. For weak coupling ($\epsilon < 0.5$) the drive and response systems remain independent, and the shape of the histogram reflects the structures in the probability density function (PDF) of the two chaotic attractors, projected along the observed variable; for strong coupling ($\epsilon > 0.8$) there is high-quality synchronization, with the distance between the systems fluctuating around the origin. As we tune the coupling strength from a (completely) unsynchronized state to a highly-synchronized state, $0.5 \leq \epsilon < 0.8$, the system exhibits attractor bubbling, characterized by the occurrence of brief escapes from the state of high-quality synchronization and by a heavy-tailed distribution for $|x_{\bot}|_n$. 

In this regime of moderate coupling, the distributions are visually similar to power laws. They nearly follow a straight line in log-log scale, at least for a certain range of values of burst sizes.  Many physical systems have limits to the maximum and minimum size produced by their variables, or to the sizes that can be observed experimentally \cite{Klaus2011}, hence it is common to find truncated power laws, both in the limit of large and small observables. We used such truncated power-law distributions to fit our data in log-log scale, yielding a single free parameter: the slope of the straight line, which gives the exponent of the power law. Notice that the slope of the straight-line fits in Fig.\ \ref{fig:HistError} changes with $\epsilon$, indicating that it can be easily tuned. 
The values of the exponents obtained for different values of the coupling parameter $\epsilon$ are  listed in Table \ref{tab:exponents}. The error bars in this table only measure the uncertainty in the slope of the optimal linear fit (in log-scale), not the true error in the value of $\alpha$, which we estimate to be on the order of 15\%. We see that, for $\epsilon$ between 0.5 and 0.6 the system crosses the critical value of exponent $\alpha = 3$, which for the pure power law implies the divergence of the second moment of the distribution, although in our system the power law is truncated, because there is a maximum value of $|x_{\bot}|_n$.

\begin{table}
\caption{\label{tab:exponents} Values of power-law exponents $\alpha$ for the fits of the experimental and numerical curves shown in Fig.\ \ref{fig:HistError}.}
\begin{tabular}{c  c  c}
\hline \hline
$\epsilon$ & { } $\alpha$ (Experimental) { } & $\alpha$ (Numerical) \\
\hline
0.5 & $2.74 \pm 0.02$ & $2.44 \pm 0.02$ \\ 
0.6 & $3.70 \pm 0.05$ & $4.06 \pm 0.03$ \\ 
0.7 & $5.67 \pm 0.15$ & $5.25 \pm 0.08$ \\ 
\hline \hline
\end{tabular}
\end{table}

The variation of the power-law exponents with the coupling level is confirmed by the numerical model.
However, due to the simplicity of our model, the parameters $A$ and $f$ (respectively, amplitude and frequency of the external pumping) need to be adjusted to obtain better agreement between the experimental and numerical distributions.
As we can see from Fig.\ \ref{fig:HistError}(a), \ref{fig:HistError}(b) and Table \ref{tab:exponents}, by changing the value of $\epsilon$ in the regime of moderate coupling we are able to control the slope of the power-law distribution of the maxima series $|x_{\bot}|_{n}$ both in the experimental circuit and in the simplified numerical model. In other words, we tune the average value of $|x_{\bot}|_{n}$, and the overall range of the distribution of this variable, turning the system more stable or unstable.


\section{Conclusion}
We investigated experimentally a pair of coupled chaotic oscillators and observed the regimes of high-quality synchronization, attractor bubbling and independent chaotic oscillation. The oscillators display attractor bubbling for moderate coupling strengths. The statistics of the size of the desynchronization events shows a heavy-tailed distribution, similar to truncated power laws appearing in many complex systems. Most interesting, we are able to tune the power-law exponent via the coupling parameter and, as a consequence, to change the range and the probability of large events.
As far as we know there is no other reported system presenting such a property. The mechanisms responsible for the occurrence of large events and the variation of the power-law exponent have implications on the predictability of extreme events happening in complex systems. 
In other words, this simple system is a starting point in the search of predicting features of other, more complicated natural or artificial systems.

\begin{acknowledgments}
We thank financial support from brazilian agencies CNPq, FINEP and CAPES.
GFOJ, HLDSC, MO thank Daniel J Gauthier for useful discussions, and for pointing out the existence of bubbling in our system.
\end{acknowledgments}


\begin{thebibliography}{45}%
\makeatletter
\providecommand \@ifxundefined [1]{%
 \@ifx{#1\undefined}
}%
\providecommand \@ifnum [1]{%
 \ifnum #1\expandafter \@firstoftwo
 \else \expandafter \@secondoftwo
 \fi
}%
\providecommand \@ifx [1]{%
 \ifx #1\expandafter \@firstoftwo
 \else \expandafter \@secondoftwo
 \fi
}%
\providecommand \natexlab [1]{#1}%
\providecommand \enquote  [1]{``#1''}%
\providecommand \bibnamefont  [1]{#1}%
\providecommand \bibfnamefont [1]{#1}%
\providecommand \citenamefont [1]{#1}%
\providecommand \href@noop [0]{\@secondoftwo}%
\providecommand \href [0]{\begingroup \@sanitize@url \@href}%
\providecommand \@href[1]{\@@startlink{#1}\@@href}%
\providecommand \@@href[1]{\endgroup#1\@@endlink}%
\providecommand \@sanitize@url [0]{\catcode `\\12\catcode `\$12\catcode
  `\&12\catcode `\#12\catcode `\^12\catcode `\_12\catcode `\%12\relax}%
\providecommand \@@startlink[1]{}%
\providecommand \@@endlink[0]{}%
\providecommand \url  [0]{\begingroup\@sanitize@url \@url }%
\providecommand \@url [1]{\endgroup\@href {#1}{\urlprefix }}%
\providecommand \urlprefix  [0]{URL }%
\providecommand \Eprint [0]{\href }%
\providecommand \doibase [0]{http://dx.doi.org/}%
\providecommand \selectlanguage [0]{\@gobble}%
\providecommand \bibinfo  [0]{\@secondoftwo}%
\providecommand \bibfield  [0]{\@secondoftwo}%
\providecommand \translation [1]{[#1]}%
\providecommand \BibitemOpen [0]{}%
\providecommand \bibitemStop [0]{}%
\providecommand \bibitemNoStop [0]{.\EOS\space}%
\providecommand \EOS [0]{\spacefactor3000\relax}%
\providecommand \BibitemShut  [1]{\csname bibitem#1\endcsname}%
\let\auto@bib@innerbib\@empty
\bibitem [{\citenamefont {Pecora}\ and\ \citenamefont
  {Carroll}(1990)}]{Pecora1990}%
  \BibitemOpen
  \bibfield  {author} {\bibinfo {author} {\bibfnamefont {L.~M.}\ \bibnamefont
  {Pecora}}\ and\ \bibinfo {author} {\bibfnamefont {T.~L.}\ \bibnamefont
  {Carroll}},\ }\bibfield  {title} {\enquote {\bibinfo {title} {Synchronization
  in chaotic systems},}\ }\href@noop {} {\bibfield  {journal} {\bibinfo
  {journal} {Phys. Rev. Lett.}\ }\textbf {\bibinfo {volume} {64}},\ \bibinfo
  {pages} {821--824} (\bibinfo {year} {1990})}\BibitemShut {NoStop}%
\bibitem [{\citenamefont {Cavalcante}\ \emph {et~al.}(2013)\citenamefont
  {Cavalcante}, \citenamefont {Ori\'{a}}, \citenamefont {Sornette},
  \citenamefont {Ott},\ and\ \citenamefont {Gauthier}}]{Cavalcante2013}%
  \BibitemOpen
  \bibfield  {author} {\bibinfo {author} {\bibfnamefont {H.~L. D.~S.}\
  \bibnamefont {Cavalcante}}, \bibinfo {author} {\bibfnamefont
  {M.}~\bibnamefont {Ori\'{a}}}, \bibinfo {author} {\bibfnamefont
  {D.}~\bibnamefont {Sornette}}, \bibinfo {author} {\bibfnamefont
  {E.}~\bibnamefont {Ott}}, \ and\ \bibinfo {author} {\bibfnamefont {D.~J.}\
  \bibnamefont {Gauthier}},\ }\bibfield  {title} {\enquote {\bibinfo {title}
  {Predictability and suppression of extreme events in a chaotic system},}\
  }\href@noop {} {\bibfield  {journal} {\bibinfo  {journal} {to be published}\
  } (\bibinfo {year} {2013})}\BibitemShut {NoStop}%
\bibitem [{\citenamefont {Cuomo}\ and\ \citenamefont
  {Oppenheim}(1993)}]{cuomo1993circuit}%
  \BibitemOpen
  \bibfield  {author} {\bibinfo {author} {\bibfnamefont {K.~M.}\ \bibnamefont
  {Cuomo}}\ and\ \bibinfo {author} {\bibfnamefont {A.~V.}\ \bibnamefont
  {Oppenheim}},\ }\bibfield  {title} {\enquote {\bibinfo {title} {Circuit
  implementation of synchronized chaos with applications to communications},}\
  }\href@noop {} {\bibfield  {journal} {\bibinfo  {journal} {Phys. Rev. Lett.}\
  }\textbf {\bibinfo {volume} {71}},\ \bibinfo {pages} {65--68} (\bibinfo
  {year} {1993})}\BibitemShut {NoStop}%
\bibitem [{\citenamefont {Kim}\ \emph {et~al.}(2006)\citenamefont {Kim},
  \citenamefont {Sramek}, \citenamefont {Uchida},\ and\ \citenamefont
  {Roy}}]{kim2006synchronization}%
  \BibitemOpen
  \bibfield  {author} {\bibinfo {author} {\bibfnamefont {M.-Y.}\ \bibnamefont
  {Kim}}, \bibinfo {author} {\bibfnamefont {C.}~\bibnamefont {Sramek}},
  \bibinfo {author} {\bibfnamefont {A.}~\bibnamefont {Uchida}}, \ and\ \bibinfo
  {author} {\bibfnamefont {R.}~\bibnamefont {Roy}},\ }\bibfield  {title}
  {\enquote {\bibinfo {title} {Synchronization of unidirectionally coupled
  {M}ackey-{G}lass analog circuits with frequency bandwidth limitations},}\
  }\href@noop {} {\bibfield  {journal} {\bibinfo  {journal} {Phys. Rev. E}\
  }\textbf {\bibinfo {volume} {74}},\ \bibinfo {pages} {016211} (\bibinfo
  {year} {2006})}\BibitemShut {NoStop}%
\bibitem [{\citenamefont {Senthilkumar}\ \emph {et~al.}(2010)\citenamefont
  {Senthilkumar}, \citenamefont {Srinivasan}, \citenamefont {Murali},
  \citenamefont {Lakshmanan},\ and\ \citenamefont
  {Kurths}}]{senthilkumar2010experimental}%
  \BibitemOpen
  \bibfield  {author} {\bibinfo {author} {\bibfnamefont {D.}~\bibnamefont
  {Senthilkumar}}, \bibinfo {author} {\bibfnamefont {K.}~\bibnamefont
  {Srinivasan}}, \bibinfo {author} {\bibfnamefont {K.}~\bibnamefont {Murali}},
  \bibinfo {author} {\bibfnamefont {M.}~\bibnamefont {Lakshmanan}}, \ and\
  \bibinfo {author} {\bibfnamefont {J.}~\bibnamefont {Kurths}},\ }\bibfield
  {title} {\enquote {\bibinfo {title} {Experimental confirmation of chaotic
  phase synchronization in coupled time-delayed electronic circuits},}\
  }\href@noop {} {\bibfield  {journal} {\bibinfo  {journal} {Phys. Rev. E}\
  }\textbf {\bibinfo {volume} {82}},\ \bibinfo {pages} {065201} (\bibinfo
  {year} {2010})}\BibitemShut {NoStop}%
\bibitem [{\citenamefont {Zamora-Munt}\ \emph {et~al.}(2010)\citenamefont
  {Zamora-Munt}, \citenamefont {Masoller}, \citenamefont {Garcia-Ojalvo},\ and\
  \citenamefont {Roy}}]{zamora2010crowd}%
  \BibitemOpen
  \bibfield  {author} {\bibinfo {author} {\bibfnamefont {J.}~\bibnamefont
  {Zamora-Munt}}, \bibinfo {author} {\bibfnamefont {C.}~\bibnamefont
  {Masoller}}, \bibinfo {author} {\bibfnamefont {J.}~\bibnamefont
  {Garcia-Ojalvo}}, \ and\ \bibinfo {author} {\bibfnamefont {R.}~\bibnamefont
  {Roy}},\ }\bibfield  {title} {\enquote {\bibinfo {title} {Crowd synchrony and
  quorum sensing in delay-coupled lasers},}\ }\href@noop {} {\bibfield
  {journal} {\bibinfo  {journal} {Phys. Rev. Lett.}\ }\textbf {\bibinfo
  {volume} {105}},\ \bibinfo {pages} {264101} (\bibinfo {year}
  {2010})}\BibitemShut {NoStop}%
\bibitem [{\citenamefont {Nixon}\ \emph {et~al.}(2011)\citenamefont {Nixon},
  \citenamefont {Friedman}, \citenamefont {Ronen}, \citenamefont {Friesem},
  \citenamefont {Davidson},\ and\ \citenamefont
  {Kanter}}]{nixon2011synchronized}%
  \BibitemOpen
  \bibfield  {author} {\bibinfo {author} {\bibfnamefont {M.}~\bibnamefont
  {Nixon}}, \bibinfo {author} {\bibfnamefont {M.}~\bibnamefont {Friedman}},
  \bibinfo {author} {\bibfnamefont {E.}~\bibnamefont {Ronen}}, \bibinfo
  {author} {\bibfnamefont {A.~A.}\ \bibnamefont {Friesem}}, \bibinfo {author}
  {\bibfnamefont {N.}~\bibnamefont {Davidson}}, \ and\ \bibinfo {author}
  {\bibfnamefont {I.}~\bibnamefont {Kanter}},\ }\bibfield  {title} {\enquote
  {\bibinfo {title} {Synchronized cluster formation in coupled laser
  networks},}\ }\href@noop {} {\bibfield  {journal} {\bibinfo  {journal} {Phys.
  Rev. Lett.}\ }\textbf {\bibinfo {volume} {106}},\ \bibinfo {pages} {223901}
  (\bibinfo {year} {2011})}\BibitemShut {NoStop}%
\bibitem [{\citenamefont {DeShazer}\ \emph {et~al.}(2001)\citenamefont
  {DeShazer}, \citenamefont {Breban}, \citenamefont {Ott},\ and\ \citenamefont
  {Roy}}]{deshazer2001detecting}%
  \BibitemOpen
  \bibfield  {author} {\bibinfo {author} {\bibfnamefont {D.~J.}\ \bibnamefont
  {DeShazer}}, \bibinfo {author} {\bibfnamefont {R.}~\bibnamefont {Breban}},
  \bibinfo {author} {\bibfnamefont {E.}~\bibnamefont {Ott}}, \ and\ \bibinfo
  {author} {\bibfnamefont {R.}~\bibnamefont {Roy}},\ }\bibfield  {title}
  {\enquote {\bibinfo {title} {Detecting phase synchronization in a chaotic
  laser array},}\ }\href@noop {} {\bibfield  {journal} {\bibinfo  {journal}
  {Phys. Rev. Lett.}\ }\textbf {\bibinfo {volume} {87}},\ \bibinfo {pages}
  {044101} (\bibinfo {year} {2001})}\BibitemShut {NoStop}%
\bibitem [{\citenamefont {Maritan}\ and\ \citenamefont
  {Banavar}(1994)}]{Maritan1994}%
  \BibitemOpen
  \bibfield  {author} {\bibinfo {author} {\bibfnamefont {A.}~\bibnamefont
  {Maritan}}\ and\ \bibinfo {author} {\bibfnamefont {J.~R.}\ \bibnamefont
  {Banavar}},\ }\bibfield  {title} {\enquote {\bibinfo {title} {Chaos, noise,
  and synchronization},}\ }\href@noop {} {\bibfield  {journal} {\bibinfo
  {journal} {Phys. Rev. Lett.}\ }\textbf {\bibinfo {volume} {72}},\ \bibinfo
  {pages} {1451--1454} (\bibinfo {year} {1994})}\BibitemShut {NoStop}%
\bibitem [{\citenamefont {Pyragas}(1996)}]{Pyragas1996}%
  \BibitemOpen
  \bibfield  {author} {\bibinfo {author} {\bibfnamefont {K.}~\bibnamefont
  {Pyragas}},\ }\bibfield  {title} {\enquote {\bibinfo {title} {Weak and strong
  synchronization of chaos},}\ }\href@noop {} {\bibfield  {journal} {\bibinfo
  {journal} {Phys. Rev. E}\ }\textbf {\bibinfo {volume} {54}},\ \bibinfo
  {pages} {R4508--R4511} (\bibinfo {year} {1996})}\BibitemShut {NoStop}%
\bibitem [{\citenamefont {Herzel}\ and\ \citenamefont
  {Freund}(1995)}]{Herzel1995}%
  \BibitemOpen
  \bibfield  {author} {\bibinfo {author} {\bibfnamefont {H.}~\bibnamefont
  {Herzel}}\ and\ \bibinfo {author} {\bibfnamefont {J.}~\bibnamefont
  {Freund}},\ }\bibfield  {title} {\enquote {\bibinfo {title} {Chaos, noise,
  and synchronization reconsidered},}\ }\href@noop {} {\bibfield  {journal}
  {\bibinfo  {journal} {Phys. Rev. E}\ }\textbf {\bibinfo {volume} {52}},\
  \bibinfo {pages} {3238--3241} (\bibinfo {year} {1995})}\BibitemShut {NoStop}%
\bibitem [{\citenamefont {Masoller}, \citenamefont {Cavalcante},\ and\
  \citenamefont {Rios~Leite}(2001)}]{masoller2001delayed}%
  \BibitemOpen
  \bibfield  {author} {\bibinfo {author} {\bibfnamefont {C.}~\bibnamefont
  {Masoller}}, \bibinfo {author} {\bibfnamefont {H.~L. D.~S.}\ \bibnamefont
  {Cavalcante}}, \ and\ \bibinfo {author} {\bibfnamefont {J.~R.}\ \bibnamefont
  {Rios~Leite}},\ }\bibfield  {title} {\enquote {\bibinfo {title} {Delayed
  coupling of logistic maps},}\ }\href@noop {} {\bibfield  {journal} {\bibinfo
  {journal} {Phys. Rev. E}\ }\textbf {\bibinfo {volume} {64}},\ \bibinfo
  {pages} {037202} (\bibinfo {year} {2001})}\BibitemShut {NoStop}%
\bibitem [{\citenamefont {Acebr{\'o}n}\ \emph {et~al.}(2005)\citenamefont
  {Acebr{\'o}n}, \citenamefont {Bonilla}, \citenamefont {Vicente},
  \citenamefont {Ritort},\ and\ \citenamefont {Spigler}}]{Acebron2005}%
  \BibitemOpen
  \bibfield  {author} {\bibinfo {author} {\bibfnamefont {J.~A.}\ \bibnamefont
  {Acebr{\'o}n}}, \bibinfo {author} {\bibfnamefont {L.~L.}\ \bibnamefont
  {Bonilla}}, \bibinfo {author} {\bibfnamefont {C.~J.~P.}\ \bibnamefont
  {Vicente}}, \bibinfo {author} {\bibfnamefont {F.}~\bibnamefont {Ritort}}, \
  and\ \bibinfo {author} {\bibfnamefont {R.}~\bibnamefont {Spigler}},\
  }\bibfield  {title} {\enquote {\bibinfo {title} {The {K}uramoto model: A
  simpler paradigm for synchronization phenomema},}\ }\href@noop {} {\bibfield
  {journal} {\bibinfo  {journal} {Rev. Mod. Phys.}\ }\textbf {\bibinfo {volume}
  {77}},\ \bibinfo {pages} {137--185} (\bibinfo {year} {2005})}\BibitemShut
  {NoStop}%
\bibitem [{\citenamefont {Boccaletti}\ \emph {et~al.}(2002)\citenamefont
  {Boccaletti}, \citenamefont {Kurths}, \citenamefont {Osipov}, \citenamefont
  {Valladares},\ and\ \citenamefont {Zhou}}]{boccaletti2002synchronization}%
  \BibitemOpen
  \bibfield  {author} {\bibinfo {author} {\bibfnamefont {S.}~\bibnamefont
  {Boccaletti}}, \bibinfo {author} {\bibfnamefont {J.}~\bibnamefont {Kurths}},
  \bibinfo {author} {\bibfnamefont {G.}~\bibnamefont {Osipov}}, \bibinfo
  {author} {\bibfnamefont {D.}~\bibnamefont {Valladares}}, \ and\ \bibinfo
  {author} {\bibfnamefont {C.}~\bibnamefont {Zhou}},\ }\bibfield  {title}
  {\enquote {\bibinfo {title} {The synchronization of chaotic systems},}\
  }\href@noop {} {\bibfield  {journal} {\bibinfo  {journal} {Physics Reports}\
  }\textbf {\bibinfo {volume} {366}},\ \bibinfo {pages} {1--101} (\bibinfo
  {year} {2002})}\BibitemShut {NoStop}%
\bibitem [{\citenamefont {Rosenblum}, \citenamefont {Pikovsky},\ and\
  \citenamefont {Kurths}(1996)}]{Rosenblum1996}%
  \BibitemOpen
  \bibfield  {author} {\bibinfo {author} {\bibfnamefont {M.~G.}\ \bibnamefont
  {Rosenblum}}, \bibinfo {author} {\bibfnamefont {A.~S.}\ \bibnamefont
  {Pikovsky}}, \ and\ \bibinfo {author} {\bibfnamefont {J.}~\bibnamefont
  {Kurths}},\ }\bibfield  {title} {\enquote {\bibinfo {title} {Phase
  synchronization of chaotic oscillators},}\ }\href@noop {} {\bibfield
  {journal} {\bibinfo  {journal} {Phys. Rev. Lett.}\ }\textbf {\bibinfo
  {volume} {76}},\ \bibinfo {pages} {1804--1807} (\bibinfo {year}
  {1996})}\BibitemShut {NoStop}%
\bibitem [{\citenamefont {Rulkov}, \citenamefont {Sushchik},\ and\
  \citenamefont {Tsimring}(1995)}]{Rulkov1995}%
  \BibitemOpen
  \bibfield  {author} {\bibinfo {author} {\bibfnamefont {N.~F.}\ \bibnamefont
  {Rulkov}}, \bibinfo {author} {\bibfnamefont {M.~M.}\ \bibnamefont
  {Sushchik}}, \ and\ \bibinfo {author} {\bibfnamefont {L.~S.}\ \bibnamefont
  {Tsimring}},\ }\bibfield  {title} {\enquote {\bibinfo {title} {Generalized
  synchronization of chaos in directionally coupled chaotic systems},}\
  }\href@noop {} {\bibfield  {journal} {\bibinfo  {journal} {Phys. Rev. E}\
  }\textbf {\bibinfo {volume} {51}},\ \bibinfo {pages} {980--994} (\bibinfo
  {year} {1995})}\BibitemShut {NoStop}%
\bibitem [{\citenamefont {Kocarev}\ and\ \citenamefont
  {Parlitz}(1996)}]{Kocarev1996}%
  \BibitemOpen
  \bibfield  {author} {\bibinfo {author} {\bibfnamefont {L.}~\bibnamefont
  {Kocarev}}\ and\ \bibinfo {author} {\bibfnamefont {U.}~\bibnamefont
  {Parlitz}},\ }\bibfield  {title} {\enquote {\bibinfo {title} {Generalized
  synchronization, predictability, and equivalence of unidirectionally coupled
  dynamical systems},}\ }\href@noop {} {\bibfield  {journal} {\bibinfo
  {journal} {Phys. Rev. Lett.}\ }\textbf {\bibinfo {volume} {76}},\ \bibinfo
  {pages} {1816--1819} (\bibinfo {year} {1996})}\BibitemShut {NoStop}%
\bibitem [{\citenamefont {Alexander}\ \emph {et~al.}(1992)\citenamefont
  {Alexander}, \citenamefont {Yorke}, \citenamefont {You},\ and\ \citenamefont
  {Kan}}]{Alexander1992}%
  \BibitemOpen
  \bibfield  {author} {\bibinfo {author} {\bibfnamefont {J.~C.}\ \bibnamefont
  {Alexander}}, \bibinfo {author} {\bibfnamefont {J.~A.}\ \bibnamefont
  {Yorke}}, \bibinfo {author} {\bibfnamefont {Z.}~\bibnamefont {You}}, \ and\
  \bibinfo {author} {\bibfnamefont {I.}~\bibnamefont {Kan}},\ }\bibfield
  {title} {\enquote {\bibinfo {title} {Riddled basins},}\ }\href@noop {}
  {\bibfield  {journal} {\bibinfo  {journal} {Intl. J. Bifurcat. Chaos}\
  }\textbf {\bibinfo {volume} {2}},\ \bibinfo {pages} {795} (\bibinfo {year}
  {1992})}\BibitemShut {NoStop}%
\bibitem [{\citenamefont {Ashwin}, \citenamefont {Buescu},\ and\ \citenamefont
  {Stewart}(1994)}]{ashwin1994bubbling}%
  \BibitemOpen
  \bibfield  {author} {\bibinfo {author} {\bibfnamefont {P.}~\bibnamefont
  {Ashwin}}, \bibinfo {author} {\bibfnamefont {J.}~\bibnamefont {Buescu}}, \
  and\ \bibinfo {author} {\bibfnamefont {I.}~\bibnamefont {Stewart}},\
  }\bibfield  {title} {\enquote {\bibinfo {title} {Bubbling of attractors and
  synchronisation of chaotic oscillators},}\ }\href@noop {} {\bibfield
  {journal} {\bibinfo  {journal} {Phys. Lett. A}\ }\textbf {\bibinfo {volume}
  {193}},\ \bibinfo {pages} {126--139} (\bibinfo {year} {1994})}\BibitemShut
  {NoStop}%
\bibitem [{\citenamefont {Maistrenko}, \citenamefont {Maistrenko},\ and\
  \citenamefont {Popovich}(1998)}]{Maistrenko1998}%
  \BibitemOpen
  \bibfield  {author} {\bibinfo {author} {\bibfnamefont {Y.~L.}\ \bibnamefont
  {Maistrenko}}, \bibinfo {author} {\bibfnamefont {V.~L.}\ \bibnamefont
  {Maistrenko}}, \ and\ \bibinfo {author} {\bibfnamefont {A.}~\bibnamefont
  {Popovich}},\ }\bibfield  {title} {\enquote {\bibinfo {title} {Transverse
  instability and riddled basins in a system of two coupled logistic maps},}\
  }\href@noop {} {\bibfield  {journal} {\bibinfo  {journal} {Phys. Rev. E}\
  }\textbf {\bibinfo {volume} {57}},\ \bibinfo {pages} {2713--2724} (\bibinfo
  {year} {1998})}\BibitemShut {NoStop}%
\bibitem [{\citenamefont {Heagy}, \citenamefont {Carroll},\ and\ \citenamefont
  {Pecora}(1995)}]{heagy1995desynchronization}%
  \BibitemOpen
  \bibfield  {author} {\bibinfo {author} {\bibfnamefont {J.~F.}\ \bibnamefont
  {Heagy}}, \bibinfo {author} {\bibfnamefont {T.~L.}\ \bibnamefont {Carroll}},
  \ and\ \bibinfo {author} {\bibfnamefont {L.~M.}\ \bibnamefont {Pecora}},\
  }\bibfield  {title} {\enquote {\bibinfo {title} {Desynchronization by
  periodic orbits},}\ }\href@noop {} {\bibfield  {journal} {\bibinfo  {journal}
  {Phys. Rev. E}\ }\textbf {\bibinfo {volume} {52}},\ \bibinfo {pages}
  {R1253--R1256} (\bibinfo {year} {1995})}\BibitemShut {NoStop}%
\bibitem [{\citenamefont {Ashwin}, \citenamefont {Buescu},\ and\ \citenamefont
  {Stewart}(1996)}]{ashwin1996attractor}%
  \BibitemOpen
  \bibfield  {author} {\bibinfo {author} {\bibfnamefont {P.}~\bibnamefont
  {Ashwin}}, \bibinfo {author} {\bibfnamefont {J.}~\bibnamefont {Buescu}}, \
  and\ \bibinfo {author} {\bibfnamefont {I.}~\bibnamefont {Stewart}},\
  }\bibfield  {title} {\enquote {\bibinfo {title} {From attractor to chaotic
  saddle: a tale of transverse instability},}\ }\href@noop {} {\bibfield
  {journal} {\bibinfo  {journal} {Nonlinearity}\ }\textbf {\bibinfo {volume}
  {9}},\ \bibinfo {pages} {703} (\bibinfo {year} {1996})}\BibitemShut {NoStop}%
\bibitem [{\citenamefont {Gauthier}\ and\ \citenamefont
  {Bienfang}(1996)}]{gauthier1996intermittent}%
  \BibitemOpen
  \bibfield  {author} {\bibinfo {author} {\bibfnamefont {D.~J.}\ \bibnamefont
  {Gauthier}}\ and\ \bibinfo {author} {\bibfnamefont {J.~C.}\ \bibnamefont
  {Bienfang}},\ }\bibfield  {title} {\enquote {\bibinfo {title} {Intermittent
  loss of synchronization in coupled chaotic oscillators: Toward a new
  criterion for high-quality synchronization},}\ }\href@noop {} {\bibfield
  {journal} {\bibinfo  {journal} {Phys. Rev. Lett.}\ }\textbf {\bibinfo
  {volume} {77}},\ \bibinfo {pages} {1751--1754} (\bibinfo {year}
  {1996})}\BibitemShut {NoStop}%
\bibitem [{\citenamefont {Venkataramani}\ \emph {et~al.}(1996)\citenamefont
  {Venkataramani}, \citenamefont {Hunt}, \citenamefont {Ott}, \citenamefont
  {Gauthier},\ and\ \citenamefont {Bienfang}}]{venkataramani1996transitions}%
  \BibitemOpen
  \bibfield  {author} {\bibinfo {author} {\bibfnamefont {S.~C.}\ \bibnamefont
  {Venkataramani}}, \bibinfo {author} {\bibfnamefont {B.~R.}\ \bibnamefont
  {Hunt}}, \bibinfo {author} {\bibfnamefont {E.}~\bibnamefont {Ott}}, \bibinfo
  {author} {\bibfnamefont {D.~J.}\ \bibnamefont {Gauthier}}, \ and\ \bibinfo
  {author} {\bibfnamefont {J.~C.}\ \bibnamefont {Bienfang}},\ }\bibfield
  {title} {\enquote {\bibinfo {title} {Transitions to bubbling of chaotic
  systems},}\ }\href@noop {} {\bibfield  {journal} {\bibinfo  {journal} {Phys.
  Rev. Lett.}\ }\textbf {\bibinfo {volume} {77}},\ \bibinfo {pages}
  {5361--5364} (\bibinfo {year} {1996})}\BibitemShut {NoStop}%
\bibitem [{\citenamefont {Krawiecki}, \citenamefont {Ho{\l}yst},\ and\
  \citenamefont {Helbing}(2002)}]{krawiecki2002volatility}%
  \BibitemOpen
  \bibfield  {author} {\bibinfo {author} {\bibfnamefont {A.}~\bibnamefont
  {Krawiecki}}, \bibinfo {author} {\bibfnamefont {J.}~\bibnamefont
  {Ho{\l}yst}}, \ and\ \bibinfo {author} {\bibfnamefont {D.}~\bibnamefont
  {Helbing}},\ }\bibfield  {title} {\enquote {\bibinfo {title} {Volatility
  clustering and scaling for financial time series due to attractor
  bubbling},}\ }\href@noop {} {\bibfield  {journal} {\bibinfo  {journal} {Phys.
  Rev. Lett.}\ }\textbf {\bibinfo {volume} {89}},\ \bibinfo {pages} {158701}
  (\bibinfo {year} {2002})}\BibitemShut {NoStop}%
\bibitem [{\citenamefont {Flunkert}\ \emph {et~al.}(2009)\citenamefont
  {Flunkert}, \citenamefont {D'Huys}, \citenamefont {Danckaert}, \citenamefont
  {Fischer},\ and\ \citenamefont {Sch{\"o}ll}}]{flunkert2009bubbling}%
  \BibitemOpen
  \bibfield  {author} {\bibinfo {author} {\bibfnamefont {V.}~\bibnamefont
  {Flunkert}}, \bibinfo {author} {\bibfnamefont {O.}~\bibnamefont {D'Huys}},
  \bibinfo {author} {\bibfnamefont {J.}~\bibnamefont {Danckaert}}, \bibinfo
  {author} {\bibfnamefont {I.}~\bibnamefont {Fischer}}, \ and\ \bibinfo
  {author} {\bibfnamefont {E.}~\bibnamefont {Sch{\"o}ll}},\ }\bibfield  {title}
  {\enquote {\bibinfo {title} {Bubbling in delay-coupled lasers},}\ }\href@noop
  {} {\bibfield  {journal} {\bibinfo  {journal} {Phys. Rev. E}\ }\textbf
  {\bibinfo {volume} {79}},\ \bibinfo {pages} {065201} (\bibinfo {year}
  {2009})}\BibitemShut {NoStop}%
\bibitem [{\citenamefont {Krawiecki}(2009)}]{krawiecki2009microscopic}%
  \BibitemOpen
  \bibfield  {author} {\bibinfo {author} {\bibfnamefont {A.}~\bibnamefont
  {Krawiecki}},\ }\bibfield  {title} {\enquote {\bibinfo {title} {Microscopic
  spin model for the stock market with attractor bubbling on scale-free
  networks},}\ }\href@noop {} {\bibfield  {journal} {\bibinfo  {journal} {J.
  Econ. Interac. Coord.}\ }\textbf {\bibinfo {volume} {4}},\ \bibinfo {pages}
  {213--220} (\bibinfo {year} {2009})}\BibitemShut {NoStop}%
\bibitem [{\citenamefont {Bak}(1999)}]{Bak1999}%
  \BibitemOpen
  \bibfield  {author} {\bibinfo {author} {\bibfnamefont {P.}~\bibnamefont
  {Bak}},\ }\href@noop {} {\emph {\bibinfo {title} {How nature works: the
  science of self-organized criticality}}}\ (\bibinfo  {publisher}
  {Copernicus},\ \bibinfo {address} {New York},\ \bibinfo {year}
  {1999})\BibitemShut {NoStop}%
\bibitem [{\citenamefont {Bak}, \citenamefont {Tang},\ and\ \citenamefont
  {Wiesenfeld}(1987)}]{Bak1987}%
  \BibitemOpen
  \bibfield  {author} {\bibinfo {author} {\bibfnamefont {P.}~\bibnamefont
  {Bak}}, \bibinfo {author} {\bibfnamefont {C.}~\bibnamefont {Tang}}, \ and\
  \bibinfo {author} {\bibfnamefont {K.}~\bibnamefont {Wiesenfeld}},\ }\bibfield
   {title} {\enquote {\bibinfo {title} {Self-organized criticality: An
  explanation of the $1/f$ noise},}\ }\href@noop {} {\bibfield  {journal}
  {\bibinfo  {journal} {Phys. Rev. Lett.}\ }\textbf {\bibinfo {volume} {59}},\
  \bibinfo {pages} {381--384} (\bibinfo {year} {1987})}\BibitemShut {NoStop}%
\bibitem [{\citenamefont {White}\ and\ \citenamefont
  {Fornberg}(1998)}]{White1998}%
  \BibitemOpen
  \bibfield  {author} {\bibinfo {author} {\bibfnamefont {B.~S.}\ \bibnamefont
  {White}}\ and\ \bibinfo {author} {\bibfnamefont {B.}~\bibnamefont
  {Fornberg}},\ }\bibfield  {title} {\enquote {\bibinfo {title} {On the chance
  of freak waves at sea},}\ }\href@noop {} {\bibfield  {journal} {\bibinfo
  {journal} {J. of Fluid Mech.}\ }\textbf {\bibinfo {volume} {355}},\ \bibinfo
  {pages} {113--138} (\bibinfo {year} {1998})}\BibitemShut {NoStop}%
\bibitem [{\citenamefont {Solli}\ \emph {et~al.}(2007)\citenamefont {Solli},
  \citenamefont {Ropers}, \citenamefont {Koonath},\ and\ \citenamefont
  {Jalali}}]{Solli2007}%
  \BibitemOpen
  \bibfield  {author} {\bibinfo {author} {\bibfnamefont {D.~R.}\ \bibnamefont
  {Solli}}, \bibinfo {author} {\bibfnamefont {C.}~\bibnamefont {Ropers}},
  \bibinfo {author} {\bibfnamefont {P.}~\bibnamefont {Koonath}}, \ and\
  \bibinfo {author} {\bibfnamefont {B.}~\bibnamefont {Jalali}},\ }\bibfield
  {title} {\enquote {\bibinfo {title} {Optical rogue waves},}\ }\href@noop {}
  {\bibfield  {journal} {\bibinfo  {journal} {Nature}\ }\textbf {\bibinfo
  {volume} {450}},\ \bibinfo {pages} {1054--1057} (\bibinfo {year}
  {2007})}\BibitemShut {NoStop}%
\bibitem [{\citenamefont {Bonatto}\ \emph {et~al.}(2011)\citenamefont
  {Bonatto}, \citenamefont {Feyereisen}, \citenamefont {Barland} \emph
  {et~al.}}]{Bonatto2011}%
  \BibitemOpen
  \bibfield  {author} {\bibinfo {author} {\bibfnamefont {C.}~\bibnamefont
  {Bonatto}}, \bibinfo {author} {\bibfnamefont {M.}~\bibnamefont {Feyereisen}},
  \bibinfo {author} {\bibfnamefont {S.}~\bibnamefont {Barland}},  \emph
  {et~al.},\ }\bibfield  {title} {\enquote {\bibinfo {title} {Deterministic
  optical rogue waves},}\ }\href@noop {} {\bibfield  {journal} {\bibinfo
  {journal} {Phys. Rev. Lett.}\ }\textbf {\bibinfo {volume} {107}},\ \bibinfo
  {pages} {053901} (\bibinfo {year} {2011})}\BibitemShut {NoStop}%
\bibitem [{\citenamefont {Estoup}(1916)}]{estoup1916gammes}%
  \BibitemOpen
  \bibfield  {author} {\bibinfo {author} {\bibfnamefont {J.-B.}\ \bibnamefont
  {Estoup}},\ }\href@noop {} {\emph {\bibinfo {title} {Gammes
  St{\'e}nographiques}}}\ (\bibinfo  {publisher} {Institut St{\'e}nographique
  de France},\ \bibinfo {address} {Paris},\ \bibinfo {year} {1916})\BibitemShut
  {NoStop}%
\bibitem [{\citenamefont {Willis}\ and\ \citenamefont
  {Yule}(1922)}]{willis1922some}%
  \BibitemOpen
  \bibfield  {author} {\bibinfo {author} {\bibfnamefont {J.}~\bibnamefont
  {Willis}}\ and\ \bibinfo {author} {\bibfnamefont {G.~U.}\ \bibnamefont
  {Yule}},\ }\bibfield  {title} {\enquote {\bibinfo {title} {Some statistics of
  evolution and geographical distribution in plants and animals, and their
  significance},}\ }\href@noop {} {\bibfield  {journal} {\bibinfo  {journal}
  {Nature}\ }\textbf {\bibinfo {volume} {109}},\ \bibinfo {pages} {177--179}
  (\bibinfo {year} {1922})}\BibitemShut {NoStop}%
\bibitem [{\citenamefont {Mercadier}\ \emph {et~al.}(2009)\citenamefont
  {Mercadier}, \citenamefont {Guerin}, \citenamefont {Chevrollier},\ and\
  \citenamefont {Kaiser}}]{mercadier2009levy}%
  \BibitemOpen
  \bibfield  {author} {\bibinfo {author} {\bibfnamefont {N.}~\bibnamefont
  {Mercadier}}, \bibinfo {author} {\bibfnamefont {W.}~\bibnamefont {Guerin}},
  \bibinfo {author} {\bibfnamefont {M.}~\bibnamefont {Chevrollier}}, \ and\
  \bibinfo {author} {\bibfnamefont {R.}~\bibnamefont {Kaiser}},\ }\bibfield
  {title} {\enquote {\bibinfo {title} {L{\'e}vy flights of photons in hot
  atomic vapours},}\ }\href@noop {} {\bibfield  {journal} {\bibinfo  {journal}
  {Nature Physics}\ }\textbf {\bibinfo {volume} {5}},\ \bibinfo {pages}
  {602--605} (\bibinfo {year} {2009})}\BibitemShut {NoStop}%
\bibitem [{\citenamefont {Newman}(2005)}]{newman2005power}%
  \BibitemOpen
  \bibfield  {author} {\bibinfo {author} {\bibfnamefont {M.~E.}\ \bibnamefont
  {Newman}},\ }\bibfield  {title} {\enquote {\bibinfo {title} {Power laws,
  pareto distributions and {Z}ipf's law},}\ }\href@noop {} {\bibfield
  {journal} {\bibinfo  {journal} {Contemp. Phys.}\ }\textbf {\bibinfo {volume}
  {46}},\ \bibinfo {pages} {323--351} (\bibinfo {year} {2005})}\BibitemShut
  {NoStop}%
\bibitem [{\citenamefont {Sornette}(1998)}]{sornette1998multiplicative}%
  \BibitemOpen
  \bibfield  {author} {\bibinfo {author} {\bibfnamefont {D.}~\bibnamefont
  {Sornette}},\ }\bibfield  {title} {\enquote {\bibinfo {title} {Multiplicative
  processes and power laws},}\ }\href@noop {} {\bibfield  {journal} {\bibinfo
  {journal} {Phys. Rev. E}\ }\textbf {\bibinfo {volume} {57}},\ \bibinfo
  {pages} {4811} (\bibinfo {year} {1998})}\BibitemShut {NoStop}%
\bibitem [{\citenamefont {Gon{\c{c}}alves}\ and\ \citenamefont
  {Neto}(2011)}]{goncalves2011electrical}%
  \BibitemOpen
  \bibfield  {author} {\bibinfo {author} {\bibfnamefont {C.}~\bibnamefont
  {Gon{\c{c}}alves}}\ and\ \bibinfo {author} {\bibfnamefont {L.}~\bibnamefont
  {Neto}},\ }\bibfield  {title} {\enquote {\bibinfo {title} {Electrical
  implementation of a complete synchronization dynamic system},}\ }in\
  \href@noop {} {\emph {\bibinfo {booktitle} {Journal of Physics: Conference
  Series}}},\ Vol.\ \bibinfo {volume} {285},\ \bibinfo {organization} {IOP}\
  (\bibinfo  {publisher} {IOP Publishing},\ \bibinfo {year} {2011})\ p.\
  \bibinfo {pages} {012013}\BibitemShut {NoStop}%
\bibitem [{\citenamefont {L\'evy}(1937)}]{Levy1937}%
  \BibitemOpen
  \bibfield  {author} {\bibinfo {author} {\bibfnamefont {P.}~\bibnamefont
  {L\'evy}},\ }\href@noop {} {\emph {\bibinfo {title} {Th\'eorie de l'addition
  des variables al\'eatoires}}}\ (\bibinfo  {publisher} {Gauthier-Villars},\
  \bibinfo {address} {Paris},\ \bibinfo {year} {1937})\BibitemShut {NoStop}%
\bibitem [{\citenamefont {Mandelbrot}(1982)}]{Mandelbrot1982}%
  \BibitemOpen
  \bibfield  {author} {\bibinfo {author} {\bibfnamefont {B.~B.}\ \bibnamefont
  {Mandelbrot}},\ }\href@noop {} {\emph {\bibinfo {title} {The Fractal Geometry
  of Nature}}}\ (\bibinfo  {publisher} {Freeman},\ \bibinfo {address} {New
  York},\ \bibinfo {year} {1982})\BibitemShut {NoStop}%
\bibitem [{\citenamefont {Shlesinger}, \citenamefont {Zaslavsky},\ and\
  \citenamefont {Klafter}(1993)}]{Shlesinger1993}%
  \BibitemOpen
  \bibfield  {author} {\bibinfo {author} {\bibfnamefont {M.~F.}\ \bibnamefont
  {Shlesinger}}, \bibinfo {author} {\bibfnamefont {G.~M.}\ \bibnamefont
  {Zaslavsky}}, \ and\ \bibinfo {author} {\bibfnamefont {J.}~\bibnamefont
  {Klafter}},\ }\bibfield  {title} {\enquote {\bibinfo {title} {Strange
  kinetics},}\ }\href@noop {} {\bibfield  {journal} {\bibinfo  {journal}
  {Nature}\ }\textbf {\bibinfo {volume} {363}},\ \bibinfo {pages} {31--37}
  (\bibinfo {year} {1993})}\BibitemShut {NoStop}%
\bibitem [{\citenamefont {Strogatz}(1994)}]{Strogatz1994}%
  \BibitemOpen
  \bibfield  {author} {\bibinfo {author} {\bibfnamefont {S.}~\bibnamefont
  {Strogatz}},\ }\href@noop {} {\emph {\bibinfo {title} {Nonlinear Dynamics and
  Chaos}}}\ (\bibinfo  {publisher} {Perseus Books},\ \bibinfo {address}
  {Cambridge, Massachusetts},\ \bibinfo {year} {1994})\BibitemShut {NoStop}%
\bibitem [{Note1()}]{Note1}%
  \BibitemOpen
  \bibinfo {note} {In the L-1 norm the distance between two arbitrary points,
  $\protect \mathaccentV {vec}17E{x}_a=(x_{a1},x_{a2},x_{a3})$ and $\protect
  \mathaccentV {vec}17E{x}_b=(x_{b1},x_{b2},x_{b3})$, is given by $d =
  |x_{a1}-x_{b1}|+|x_{a2}-x_{b2}|+|x_{a3}-x_{b3}|$, instead of the square root
  of the sum of the squared differences, usual in the L-2 norm.}\BibitemShut
  {Stop}%
\bibitem [{\citenamefont {Klaus}, \citenamefont {Yu},\ and\ \citenamefont
  {Plenz}(2011)}]{Klaus2011}%
  \BibitemOpen
  \bibfield  {author} {\bibinfo {author} {\bibfnamefont {A.}~\bibnamefont
  {Klaus}}, \bibinfo {author} {\bibfnamefont {S.}~\bibnamefont {Yu}}, \ and\
  \bibinfo {author} {\bibfnamefont {D.}~\bibnamefont {Plenz}},\ }\bibfield
  {title} {\enquote {\bibinfo {title} {Statistical analyses support power law
  distributions found in neuronal avalanches},}\ }\href@noop {} {\bibfield
  {journal} {\bibinfo  {journal} {PLoS One}\ }\textbf {\bibinfo {volume} {6}},\
  \bibinfo {pages} {e19779} (\bibinfo {year} {2011})}\BibitemShut {NoStop}%
\bibitem [{\citenamefont {Christensen}\ \emph {et~al.}(2002)\citenamefont
  {Christensen}, \citenamefont {Danon}, \citenamefont {Scanlon},\ and\
  \citenamefont {Bak}}]{Christensen2002}%
  \BibitemOpen
  \bibfield  {author} {\bibinfo {author} {\bibfnamefont {K.}~\bibnamefont
  {Christensen}}, \bibinfo {author} {\bibfnamefont {L.}~\bibnamefont {Danon}},
  \bibinfo {author} {\bibfnamefont {T.}~\bibnamefont {Scanlon}}, \ and\
  \bibinfo {author} {\bibfnamefont {P.}~\bibnamefont {Bak}},\ }\bibfield
  {title} {\enquote {\bibinfo {title} {Unified scaling law for earthquakes},}\
  }\href {\doibase 10.1073/pnas.012581099} {\bibfield  {journal} {\bibinfo
  {journal} {Proc. Natl. Acad. Sci.}\ }\textbf {\bibinfo {volume} {99}},\
  \bibinfo {pages} {2509--2513} (\bibinfo {year} {2002})}\BibitemShut {NoStop}%
\end{thebibliography}
%

\end{document}